\documentclass[sigconf,nonacm]{acmart} % ICS

\setcopyright{none}

\usepackage[normalem]{ulem}

\usepackage[T1]{fontenc}

\usepackage{ifthen}
\usepackage{import}
\usepackage{textcomp}
\usepackage[vlined,ruled]{algorithm2e}
\usepackage{verbatim}
\usepackage{textgreek}
\usepackage{tikz}
\usetikzlibrary{matrix, plotmarks, chains, positioning, shapes.multipart, calc, shapes, arrows, shapes.arrows}
\usetikzlibrary{calc}
\usepackage{commath}
\usepackage{multirow}
\usepackage{mdframed}
\usepackage{booktabs}
\usepackage{makecell}

\usepackage{pgf}
\usepackage{pgfplots}
\pgfplotsset{compat=1.16}
\usepackage[super]{nth}

\usepackage{subcaption}

\usepackage{verbatim}
\usepackage{paralist}

\usepackage{url}

\usepackage[shortcuts]{extdash}

\newcommand{\void}[1]{}

\usepackage{pdfcomment}
\newcommand{\todo}[1]{\marginpar{\pdfcomment[color=red]{Todo: #1}}}
\newcommand{\reftab}[1]{Table~\ref{tab:#1}}
\newcommand{\reffig}[1]{Figure~\ref{fig:#1}}

\newcommand{\refsec}[1]{Section~\ref{sec:#1}}

\newcommand{\labeltab}[1]{\label{tab:#1}}

\newcommand{\labfig}[1]{\label{fig:#1}}

\newcommand{\labsec}[1]{\label{sec:#1}}

\usepackage{xspace}

\newlength{\tmpTFS}

\newcommand{\microops}{{\textmu}ops\xspace}
\newcommand{\microop}{{\textmu}op\xspace}

\newcommand{\bhivel}{\emph{BHive\textsubscript{L}}\xspace}
\newcommand{\bhiveu}{\emph{BHive\textsubscript{U}}\xspace}

\newcommand{\uiCA}{uiCA\xspace}

\makeatletter
\let\zz\Gscale@box
\long\def\Gscale@box#1{%
\xdef\thelastscalefactor{#1}%
\zz{#1}}
\makeatother

\let\oldFootnote\footnote
\newcommand\nextToken\relax
\renewcommand\footnote[1]{%
    \oldFootnote{#1}\futurelet\nextToken\isFootnote}
\newcommand\isFootnote{%
    \ifx\footnote\nextToken\textsuperscript{,}\fi}								  

\begin{document}

\title[uiCA: Accurate Throughput Prediction of Basic Blocks on Recent Intel Microarchitectures]{uiCA: Accurate Throughput Prediction of\\Basic Blocks on Recent Intel Microarchitectures}
%\title[A Parametric Microarchitecture Model for Accurate Basic Block Throughput Prediction on Recent Intel CPUs]{A Parametric Microarchitecture Model for Accurate\\Basic Block Throughput Prediction on Recent Intel CPUs}

\author{Andreas Abel and Jan Reineke}
\email{{abel, reineke}@cs.uni-saarland.de}
\affiliation{%
  \institution{Saarland University\\
  Saarland Informatics Campus}
  \city{Saarbr\"ucken}
  \country{Germany}
}

\begin{abstract} 
Performance models that statically predict the steady-state throughput of basic blocks on particular microarchitectures, such as IACA, Ithemal, llvm-mca, OSACA, or CQA, can guide optimizing compilers and aid manual software optimization.
However, their utility heavily depends on the accuracy of their predictions.
The average error of existing models compared to measurements on the actual hardware has been shown to lie between 9\% and 36\%.
But how good is this?
To answer this question, we propose an extremely simple analytical throughput model that may serve as a baseline. 
Surprisingly, this model is already competitive with the state of the art, indicating that there is significant potential for improvement.

To explore this potential, 
we develop a simulation-based throughput predictor.
To this end, we propose a detailed parametric pipeline model that supports all Intel Core microarchitecture generations released between 2011 and 2021.
We evaluate our predictor on an improved version of the BHive benchmark suite and show that its predictions are usually within 1\% of measurement results, improving upon prior models by roughly an order of magnitude.
The experimental evaluation also demonstrates that several microarchitectural details considered to be rather insignificant in previous work, are in fact essential for accurate prediction.

Our throughput predictor is available as open source.
\end{abstract}

\maketitle

\section{Introduction}

%Models to predict software behavior on hardware are ubiquitous 
Performance models are widely used to predict, understand, and optimize performance.  %hardware and software performance.
Such models have been proposed at all kinds of abstraction levels, targeting different resources and applications. 
In this work, we are considering the problem of predicting the steady-state throughput of basic blocks, i.e., the number of processor clock cycles it takes to execute a basic block in steady state in a loop.\looseness=-1

Being able to accurately predict basic-block throughput is important both for compiler designers and for performance engineers.
Compiler optimizations, for instance, may rely on such performance models to apply auto-vectorization, register allocation, and instruction scheduling~\cite{mcgovern99,stephenson03,lozano12} in an informed manner.
Performance engineers may use such models to pinpoint constraining bottlenecks, which can then be mitigated by code changes.

Clearly, the utility of such models depends greatly on the accuracy of their predictions.
Both Pohl et al.~\cite{pohl19,pohl20} and Mendis and Amarasinghe~\cite{mendis18} observe that the inaccuracy of existing cost models often misguides optimizations at the cost of performance.

In recent work, several basic block performance analysis models have been proposed, such as IACA~\cite{iacaGuide}, llvm-mca~\cite{dibiagio18,llvmmca}, OSACA~\cite{Laukemann18, Laukemann19}, CQA~\cite{cqa14}, Ithemal~\cite{mendis19a}, and DiffTune~\cite{Renda20}. 
Chen et al.~\cite{Chen19} developed the BHive benchmark suite specifically to evaluate the accuracy of these models. 
They found that the average error of the predictions of existing tools compared to measurements on the actual hardware lies between 9\% and 36\%.

At the onset of this work, we were wondering how good an accuracy in this range actually is.
To shed light on this question, we first developed the following simple analytical throughput predictor that can serve as a baseline.
We predict the throughput of a benchmark on the Skylake microarchitecture as
\[TP_{\textit{baseline}} = max\left(\frac{n}{4}, \frac{m_r}{2}, m_w\right),\]
where $n$ is the number of instructions, $m_r$ the number of memory reads, and $m_w$ the number of memory writes of the benchmark. 
Note that Skylake has four decoders, and it can perform two memory reads and one memory write per cycle.
Thus, the three constituents of the formula directly follow from basic throughput limits of the microarchitecture. % to simple throughput limits imposed by the decoder, as well as reads and writes.

\reftab{baseline} shows the mean absolute percentage error (MAPE) and Kendall's tau coefficient (for details see \refsec{evaluation}) of the predictions of different tools relative to the reference measurements that were published along with the BHive benchmark suite~\cite{Chen19} for the Skylake microarchitecture.
Surprisingly, our simple baseline predictor achieves an average error of around 17\%, which is better than several previous approaches that are significantly more complex.

%\pagebreak

\begin{table}
\caption{State of the art in basic-block throughput prediction}% for the Skylake microarchitecture}
\labeltab{baseline}
%\negvspacesmall
\begin{center}
\begin{tabular}{lrc}
\toprule
\textbf{Predictor} & \textbf{MAPE} & \textbf{Kendall's Tau} \\
\midrule
Ithemal & 9.51\% & 0.8523 \\
IACA 3.0 & 14.50\% & 0.8131 \\
DiffTune & 24.62\% & 0.7444 \\
llvm-mca-10 & 27.91\% & 0.7832 \\
         OSACA & 29.74\% & 0.7770 \\
\midrule
         Baseline & 17.21\% & 0.7719 \\
\bottomrule
\end{tabular}
\end{center}
%\negvspace
%\negvspace
\end{table}

This yields the following questions that we address in this work: 
\begin{enumerate}
  \item[\textbf{Q1}] What are the reasons for the discrepancies between the predictions of existing models and measurements?
  \item[\textbf{Q2}] How can these discrepancies be eliminated?
\end{enumerate}

We have identified two main issues that explain the discrepancies between measurements and predictions: 
\begin{enumerate}
  \item The microarchitectural models of previous tools are not detailed enough.
  \item The experimental evaluations include unsuitable benchmarks, and are partly based on biased and inaccurate throughput measurements, and on incompatible throughput definitions.
\end{enumerate}
%In this paper, we address both of these issues.

To address the first issue, we develop a pipeline model that is significantly more detailed than prior models.
It is applicable to all Intel Core microarchitectures released in the last decade.
Still, the differences between these microarchitectures can be captured by a small number of parameters.
A challenge in building such a model is that many of the relevant properties are undocumented.
We have developed microbenchmark-based techniques to reverse-engineer these properties.
Based on this model, we implement a simulator called \uiCA (``uops.info Code Analyzer'') that predicts the throughput of basic blocks (\refsec{pipelineModel}).
%In addition, it also provides further insights into how basic blocks are executed, such as the port usage for each instruction. %discuss somewhere else? i'm not sure %it's sort of lost here as an isolated sentence

To address the second issue, and to evaluate \uiCA, we first identify two variants of the throughput prediction problem~(\refsec{problem}) that capture differing assumptions in the existing models.
Then we propose several improvements to the BHive benchmark suite that enable a fairer comparison of different tools.
Finally, we develop a more accurate measurement methodology that supports both problem variants~(\refsec{benchmarks}).

Using this improved benchmark suite and measurement methodology, we compare \uiCA with the existing tools on nine different Intel Core microarchitectures.
\uiCA's predictions are usually within 1\% of the measurement results, which improves upon the state of the art by roughly an order of magnitude (\refsec{evaluation}).

%show that several previous tools do not beat baseline

%\subsubsection*{Summary of contributions}

To summarize, the main contributions of our paper are:
\begin{itemize}
\item A parametric pipeline model that is applicable to \emph{all} Intel Core microarchitectures from Sandy Bridge (2011) to Rocket Lake (2021) (\refsec{pipelineModel}).
\item A throughput predictor (\refsec{TPPredictor}) based on this parametric model that is more accurate than previous work, often by more than an order of magnitude (\refsec{evaluation}).
\item An improved set of benchmarks for validating throughput predictors, and a more accurate measurement methodology (\refsec{benchmarks}). %\todo{we can argue that this would also be helpful for machine-learning based approaches, because they are essentially garbage-in, garbage-out; e.g. in discussion at the end of the paper}
%\item A simple analytical throughput predictor that serves as a baseline in the evaluation and that outperforms several previous approaches (\refsec{baseline}).
%\todo{\item A discussion of general limitations/challenges for throughput prediction.}
\end{itemize}

\section{Related Work}

%full-system simulators~\cite{austin02,yourst07,loh09,binkert11,patel11,sanchez13} such as SimpleScalar~\cite{austin02} or gem5~\cite{binkert11}

%roofline~\cite{williams09} and kerncraft (ecm)~\cite{hammer17}

%WCET analysis~\cite{ferdinand01,li07,wilhelm08}

%compilers: 
%  instruction scheduling and register allocation~\cite{mcgovern99,stephenson03,lozano12}
%  automatic vectorization~\cite{mendis18,pohl19}

%\todo{analytical models, simulation-based models, ML models}

\subsection{Models for Performance Prediction}

Existing basic-block throughput models roughly fall into three camps: (1) simulation-based models, (2) analytical models, and (3) machine-learning-based models.

The Intel Architecture Code Analyzer (IACA)~\cite{iacaGuide} is a tool developed by Intel that can statically analyze the performance of loop kernels on several older microarchitectures.
The tool generates a report which includes throughput and port usage data of the analyzed loop kernel.
As the tool is closed source, its underlying model and prediction methodology are unknown.
In April 2019, Intel announced IACA's end of life.

The Open Source Architecture Code Analyzer (OSACA)~\cite{Laukemann18, Laukemann19} is an analytical performance prediction tool for Intel, AMD, and ARM microarchitectures.
It is based on a relatively coarse-grained model of these microarchitectures; according to~\cite{Laukemann19}, ``accurately modeling the performance characteristics of the decode, reorder buffer, register allocation/renaming, retirement and other stages, which all may limit the execution throughput and impose latency penalties, is currently out of scope for OSACA.''

The LLVM Machine Code Analyzer (llvm-mca)~\cite{llvmmca} is a simulation-based tool that predicts the performance of machine code using scheduling models available in the LLVM compiler~\cite{llvm04}.
It supports processors for which a scheduling model is available in LLVM.
llvm-mca does not model performance bottlenecks in the front end.
It also does not model techniques such as micro/macro fusion or move elimination.

Within the MAQAO~\cite{maqao} framework, two performance analysis tools have been proposed: CQA~\cite{cqa14} and UFS~\cite{Palomares16}.
The Code Quality Analyzer (CQA) is a simulation-based tool that analyzes the performance of innermost loops.
In addition to computing throughput predictions, it provides high-level code quality metrics like the vectorization ratio.
CQA uses a front-end model that is more detailed than those of most other previous tools; however, it does not model the core of the execution engine ``because of its complexity and lack of documentation.''
UFS is a throughput predictor that uses a relatively detailed model of the back end of the Sandy Bridge microarchitecture, but only a very coarse-grained model of its front end.
UFS exists only as a prototype that is not publicly available.

Ithemal~\cite{mendis19a} is a basic-block throughput predictor that is based on a deep neural network.
This neural network has been trained using training data obtained from measurements on the actual hardware. 
This data-driven approach holds the promise of alleviating the tedious modeling effort required to build conventional simulators. 
%\todo{its performance critically depends on the quality of the training data; discuss here or better later that retraining Ithemal may give better results}
Unlike most of the other tools discussed in this section, it predicts a basic block's throughput but does not provide other insights into how the code is executed, which may be useful for performance engineers.
To obtain a more interpretable model, Renda et al.\ present DiffTune~\cite{Renda20}, an approach that applies machine learning to obtain the microarchitecture-specific parameters for the x86 simulation model used by llvm-mca. 
As it is based on llvm-mca, it shares its modeling limitations.
Both Ithemal and DiffTune were trained on measurements performed with a profiling tool by Chen et al.~\cite{Chen19}.
In contrast to previous tools, this tool can automatically profile basic blocks that may perform arbitrary memory accesses.
We build upon this tool in \refsec{measurements}.

Full-system simulators~\cite{austin02,yourst07,loh09,binkert11,patel11,sanchez13} such as ZSim or gem5 can simulate the execution of entire programs, modeling the interactions between different hardware components including deep memory hierarchies and multi-core architectures. 
In contrast to basic-block throughput predictors, full-system simulators require a program's input data to drive the simulation.
In principle, the pipeline model developed in this paper could be integrated into full-system simulators to improve their accuracy.

%Worst-case execution time analysis (WCET) tools~\cite{ferdinand01,li07,wilhelm08} statically compute upper bounds on a program's execution time under any possible program input.
%A common approach in WCET analysis~\cite{wilhelm08} is to first bounds the execution time of each basic block and then to combine these bounds to bound the whole program's execution time.
%We are not aware of WCET analyses that faithfully capture processors that are as complex as the ones considered in this work.
%A challenging in transferring our results to WCET analysis is that such analyses require safe upper bounds, whereas our approach is to predict average/best-case behavior.

\subsection{Microbenchmarking}

One approach to construct detailed performance models is to reverse engineer the relevant parameters by microbenchmarking.

Agner Fog~\cite{fogTest} provides a measurement tool and a set of test scripts that generate microbenchmarks to analyze various properties of microarchitectures.
He maintains a set of tables with instruction latencies, throughputs and micro-operation breakdowns~\cite{fog21}, as well as a document with detailed descriptions of many recent microarchitectures~\cite{fog21uArch}.

nanoBench~\cite{Abel20, Abel20b} is a tool for evaluating small microbenchmarks on x86 systems using hardware performance counters.
nanoBench is used to evaluate the microbenchmarks for obtaining the latency, throughput, and port usage data that is available at uops.info~\cite{Abel19}, which we extend in this work and employ in our tool.
%which we use in our simulator?

PMEvo~\cite{Ritter20} is a framework by Ritter and Hack that can automatically infer port mappings based on measured execution times of short code sequences.
A related approach that also takes into account other limiting resources besides execution ports was recently proposed by Derumigny et al.~\cite{derumigny22}.
The models obtained by these two approaches may be used to predict the throughput of dependency-free basic blocks.

\section{Basic-Block~Throughput~Prediction}\labsec{problem}

In this section, we state more precisely the problem basic-block throughput predictors aim to solve.

\newcommand{\DefTPL}{$TP_L$\xspace}
\newcommand{\DefTPU}{$TP_U$\xspace}
\subsection{Notions of Throughput}\labsec{definitions}
The throughput of a basic block is commonly defined as the average number of clock cycles per iteration when executing the basic block repeatedly in a steady state. 

However, ``executing the basic block repeatedly'' can mean different things, depending on the type of basic block that is used.

For basic blocks that end in a branch instruction that jumps back to the beginning of the block, ``executing the basic block repeatedly'' can reasonably be interpreted to mean executing the block in a way that the branch is always taken; this corresponds to executing the basic block as an infinite loop.
In the following, we will refer to this notion of throughput as \DefTPL.

Basic blocks that do not end in a branch instruction cannot be executed in this way.
For such blocks, a way to ``execute the basic block repeatedly'' is to unroll the basic block a sufficient number of times to reach a steady state.
In the following, we will refer to this notion of throughput as \DefTPU.

An important difference between the two notions is that for \DefTPL, the \microops of many benchmarks are delivered by the µOP cache (DSB) or the loop stream detector (see \refsec{pipelineModel}), whereas for \DefTPU, all \microops have to go through the decoders, which can be significantly slower.
As an example, consider the basic block
\begin{center}
\texttt{ADD AX, 0x1234; DEC R15}
\end{center}
When unrolling this code sequence multiple times, the average execution time on a Skylake CPU is $3.44$ cycles per iteration; the bottleneck here is a stall in the predecoder.
On the other hand, the code sequence 
\begin{center}
\texttt{loop: ADD AX, 0x1234; DEC R15; JNZ loop}
\end{center}
is served from the µOP cache (DSB) and requires, on average, only one cycle per loop iteration, even though it has an additional instruction.

Previous work did not clearly distinguish the two throughput notions.
Intel's IACA treats the basic block as the ``body of an infinite loop''.
Thus, it is based on \DefTPL.
Correspondingly, all examples in IACA's user guide~\cite{iacaGuide} are basic blocks that end in a branch instruction that jumps back to the beginning of the block.
IACA does not reject basic blocks that are not of this form; however, the behavior in such a case is not specified in the documentation.
Chen et al.\ claim to use ``IACA's definition of throughput''~\cite{Chen19}.
However, their measurement framework only considers basic blocks that do not end in a branch instruction.
They measure the throughput by unrolling; thus their notion of throughput actually corresponds to~\DefTPU.%\looseness=-1

OSACA~\cite{Laukemann18, Laukemann19} and CQA~\cite{cqa14} are based on \DefTPL; CQA can only analyze code that ends in a branch instruction.
For llvm-mca, it is not completely clear which definition is used.
According to the documentation, llvm-mca ``simulates the execution of the machine code sequence in a loop of iterations''~\cite{llvmmca}; however, the examples in the documentation do not end in a branch instruction.
As llvm-mca does not model performance bottlenecks in the front end, the throughput predictions can generally be expected to be closer to measurements based on the \DefTPL notion.

In this work, we develop a throughput predictor that can predict the throughput under both notions. 
For basic blocks that end in a branch instruction, we use \DefTPL; for other blocks, we use \DefTPU.

\subsection{Common Modeling Assumptions}\labsec{assumptions}  

A common and sometimes unstated assumption is that basic-block throughput predictors are intended to statically analyze the performance of compute-bound basic blocks, i.e., basic blocks whose throughput is not memory- or I/O-bound. 
A basic-block throughput predictor may be one component of tools and methodologies to determine whether code is actually compute bound, such as the Roofline model~\cite{williams09} or the Execution-Cache-Memory model~\cite{Stengel15,hammer17}; however, determining compute-boundedness is not in scope for basic-block throughput predictors themselves.

Further modeling assumptions arise from the fact that basic blocks are analyzed statically without knowledge of the state of the execution environment, such as the values of registers, memory, or microarchitectural components such as branch predictors.
%
%Existing basic-block throughput predictors make several assumptions on the kinds of basic blocks that they are intended to analyze, as well as on the execution environment. 
%These assumptions are necessary because the analysis is performed statically, i.e, without program inputs and without knowledge of the hardware state at the onset.
%

We summarize these common modeling assumptions in the following:
\begin{compactitem}
\item All memory accesses are executed optimally. This means there are no cache or TLB misses, no unaligned loads, no bank conflicts~\cite{intelOptManual20, Jiang17}, and stores can always be paired~\cite{intelOptManual20}.
%\pagebreak
\item There are no branch mispredictions.
\item The basic blocks were emitted by a compiler or written by a reasonably competent, non-adversarial programmer. 
      Thus, they do not contain corner cases that would not occur in realistic, performance-critical code like undefined instructions, x87 floating-point stack underflows or overflows, or memory accesses to invalid addresses.
\item There are no denormal floating-point operations.
\item There are no operations that lead to exceptions, and no interrupts occur during the execution.
\item A problematic class of basic blocks for static throughput predictors are blocks with input-dependent timing. 
      This includes blocks that use variable-latency instructions, and blocks for which it depends on the inputs whether two memory accesses alias. 
      Most existing tools optimistically assume that the inputs are such that no memory aliasing occurs; in llvm-mca it can be configured via a parameter whether such memory accesses are assumed to alias or not.
      For variable-latency instructions, the behavior of existing tools is inconsistent. 
      For divisions, most tools yield pessimistic predictions; for the \verb|cpuid| instruction, on the other hand, the predictions are typically optimistic.
\end{compactitem}

It should be noted that for actual programs, these assumptions do not necessarily hold.
However, techniques to check whether the assumptions hold are mostly orthogonal to the techniques required for basic-block throughput prediction, and it makes therefore sense to separate these concerns.

\begin{comment}
\subsection{Challenges}
\begin{itemize}
  \item basic-block performance (sometimes) depends on
    \begin{itemize}
      \item inputs: input-dependent instruction timing; input-dependent memory dependencies; memory-hierarchy performance
      \item initial hardware state: caches, branch predictors, TLB
      \item (hardware components)
    \end{itemize} 
    these things are statically unknown (in particular to the predictor)\\
    in order to focus on the performance provided by the processor core, a common approach is to assume ideal behavior of the unmodeled/unknown parts, i.e. memory accesses are assumed to be cache hits, branches are assumed to be predicted correctly, etc. (exception: input-dependent instruction timing
      
  \item definitions of throughput
    \begin{itemize}
      \item unrolled throughput
      \item loop throughput
      \item discussion of differences
    \end{itemize}
\end{itemize}
\end{comment}

\newmdenv[
  topline=false,
  bottomline=false,
  rightline=false, 
  usetwoside=false, 
  leftmargin=-7pt,
  innerleftmargin=5pt,
  innerrightmargin=0pt,
  linecolor=blue,
  linewidth=2pt,  
]{newInfo}

\newmdenv[
  topline=false,
  bottomline=false,
  leftline=false,
  usetwoside=false,
  leftmargin=0pt,
  rightmargin=-7pt,
  innerleftmargin=0pt,
  innerrightmargin=5pt,
  linecolor=red,
  linewidth=2pt,  
]{parameter}

\section{A Parametric Pipeline Model}\labsec{pipelineModel}
\reffig{pipeline} shows the general structure of the pipelines of recent Intel Core CPUs.
At such a high level, all these CPUs are very similar.
For developing an accurate performance predictor, however, a more detailed model is necessary.

In this section, we describe the parametric pipeline model that we have developed.
First, in \refsec{pipelineproperties}, we describe the different pipeline components of recent Intel CPUs.
Then, in \refsec{instructionProperties}, we look at properties of how individual instructions are executed.
Finally, in \refsec{TPPredictor}, we discuss how the model is used to implement a throughput predictor.

Many of the details that are necessary to build such a detailed model are undocumented.
We have reverse-engineered these details via microbenchmarks using hardware performance counters.
For evaluating these microbenchmarks, we use nanoBench~\cite{Abel20} and an extension to nanoBench that provides cycle-by-cycle performance data, similar to Brandon Falk's ``Sushi Roll'' approach~\cite{Falk19}.
Due to space constraints, we are unable to describe these microbenchmarks in this section; we will describe those in a separate technical report.
Instead, we will only describe our findings. 
We highlight the corresponding paragraphs with a blue bar on the left.

The model presented in this section is significantly more detailed than all previous models described in the literature.
It is applicable to \emph{all} Intel Core microarchitectures from Sandy Bridge to Rocket Lake.
We found that, maybe surprisingly, the differences between these microarchitectures that are relevant for our work can be captured by a relatively small number of parameters.
We mark the paragraphs that describe these parameters with a red bar on the right.\looseness=-1

%\negvspacesmall

\subsection{Pipeline Properties}\labsec{pipelineproperties}
\subsubsection{Front End}
The predecoder fetches aligned 16-byte blocks from the instruction cache (at most one such block per cycle), and it is mainly responsible for detecting where each instruction begins.
This is not completely straightforward, as an instruction can be between 1 and 15 bytes long, and detecting its length can require inspecting several bytes of the instruction.
The predecoded instructions are inserted into the \emph{instruction queue} (IQ).

% Define the layers to draw the diagram
\pgfdeclarelayer{background}
\pgfdeclarelayer{background1}
\pgfdeclarelayer{foreground}
\pgfsetlayers{background,background1,main,foreground}
 
% Define block styles  
\tikzstyle{nodeStyle} = [draw, text width=6cm,  minimum height=1.75em, text centered]
\tikzstyle{port} = [draw, fill={rgb,255:red,135; green,220; blue,170}, text width=0.75cm, font=\fontsize{6}{7.2}\sffamily, text centered]
\tikzstyle{FU} = [draw, fill={rgb,255:red,212; green,170; blue,0}, text width=5.0em, font=\fontsize{6}{7.2}\sffamily, rotate=90, text centered]
\tikzstyle{arrow} = [draw, thick, color=black!80, font=\footnotesize\sffamily]

% Draw background
\newcommand{\background}[7]{%
  \begin{pgfonlayer}{background}
    % Left-top corner of the background rectangle
    \path (#1.west |- #2.north)+(-1,0.4) node (a1) {};
    % Right-bottom corner of the background rectangle
    \path (#3.east |- #4.south)+(+0.4,#5) node (a2) {};
    % Draw the background
    \path[fill=#6, draw=black!50]
      (a1) rectangle (a2);
    \path let \p{x}=(a1), \p{y}=($(a1)!0.5!(a2)$) in (\x{x}, \y{y})+(0.5,0) node (u1)[rotate=90]
      {#7};
  \end{pgfonlayer}}
  
\begin{figure}[t]
\centering
\begin{tikzpicture}[scale=.8,transform shape,font=\fontsize{11}{13.2}\sffamily] 
  % Draw diagram elements
  \path node (nIC) [nodeStyle, fill={rgb,255:red,249; green,177; blue,166}] {Instruction Cache};
  \path (nIC.south)+(0.0,-0.75) node (nPreDec) [nodeStyle, text width=4cm, fill={rgb,255:red,171; green,204; blue,227}] {Predecoder};  
  \path (nPreDec.south)+(0,-0.75) node (nIQ) [nodeStyle, text width=4cm, fill={rgb,255:red,198; green,233; blue,175}] {Instruction Queue (IQ)};
  \path (nIQ.south)+(-2.5,-0.75) node (nDSB) [nodeStyle, text width=1cm, fill={rgb,255:red,249; green,177; blue,166}] {DSB};
  \path (nIQ.south)+(0.0,-0.75) node (nDec) [nodeStyle, text width=2cm, fill={rgb,255:red,171; green,204; blue,227}] {Decoder};
  \path (nIQ.south)+(2.5,-0.75) node (nMS) [nodeStyle, text width=1cm, fill={rgb,255:red,171; green,204; blue,227}] {MS};
  \path (nDec.south)+(0.0,-0.75) node (nIDQ) [nodeStyle, fill={rgb,255:red,198; green,233; blue,175}] {Instruction Decode Queue (IDQ)};
  \path (nIDQ.south)+(0.0,-1) node (nRenamer) [nodeStyle, fill={rgb,255:red,171; green,204; blue,227}] {Renamer / Allocator};
  
  \path (nRenamer.south)+(0,-0.75) node (nReorder) [nodeStyle, text width=3cm, fill={rgb,255:red,198; green,233; blue,175}] {Reorder Buffer};
  \path (nReorder.south)+(0.0,-0.75) node (nRS) [nodeStyle, fill={rgb,255:red,135; green,220; blue,170}] {Scheduler};
  \path (nRS.south)+(-2.5,0.0) node[anchor=north] (nPort0) [port] {Port 0};
  \path (nRS.south)+(-1.5,0.0) node[anchor=north] (nPort1) [port] {Port 1};
  \path (nRS.south)+(-0.5,0.0) node[anchor=north] (nPort2) [port] {Port 2};
  \path (nRS.south)+(0.5,0.0) node[anchor=north] (nPort3) [port] {Port 3};
  \path (nRS.south)+(1.5,0.0) node[anchor=north] (nPort4) [port] {Port 4};
  \path (nRS.south)+(2.5,0.0) node[anchor=north] (nPort5) [port] {Port 5};
  
  \path (nPort0.south)+(0,-0.75) node[anchor=east] (nPort0FU) [FU] {ALU, V-MUL, \dots};
  \path (nPort1.south)+(0,-0.75) node[anchor=east] (nPort1FU) [FU] {ALU, V-ADD, \dots};
  \path (nPort2.south)+(0,-0.75) node[anchor=east] (nPort2FU) [FU] {Load, AGU};
  \path (nPort3.south)+(0,-0.75) node[anchor=east] (nPort3FU) [FU] {Load, AGU};
  \path (nPort4.south)+(0,-0.75) node[anchor=east] (nPort4FU) [FU] {Store Data};
  \path (nPort5.south)+(0,-0.75) node[anchor=east] (nPort5FU) [FU] {ALU, JMP, \dots};
  
  \begin{pgfonlayer}{background1}    
    \path (nPort0FU.north |- nPort0FU.east)+(-0.25,0.25) node (ee_tl) {};
    \path (nPort5FU.south |- nPort5FU.west)+(+0.25,-0.25) node (ee_br) {};
    \path[fill={rgb,255:red,95; green,211; blue,188}, draw=black!50, rounded corners] (ee_tl) rectangle (ee_br);
  \end{pgfonlayer}
  
  \path let \p{x}=(nRS.south), \p{y}=(ee_br.south) in (\x{x}, \y{y})+(0,-1.0) node (nL1D) 
    [nodeStyle, fill={rgb,255:red,249; green,177; blue,166}] {L1 Data Cache};
  \path (nL1D.south)+(0.0,-0.75) node (nL2) [nodeStyle, fill={rgb,255:red,249; green,177; blue,166}] {L2 Cache};
     
  % Draw arrows between elements  
  \draw [->, arrow] (nIC.south) -- (nPreDec.north);
  \draw [->, arrow] (nPreDec.south) -- (nIQ.north);
  \draw [->, arrow] (nIQ.south) -- (nDec.north);
  \draw [->, arrow] (nDec.south) -- (nIDQ.north);
  \draw [->, arrow] (nDec.west) -- (nDSB.east);
  \draw [->, arrow] (nIDQ.south) -- (nRenamer.north);
  
  \draw [->, arrow] (nDSB.south) -- (nDSB.south |- nIDQ.north);
  \draw [->, arrow] (nMS.south) -- (nMS.south |- nIDQ.north);
  
  \path (nRenamer.south west) -- (nRenamer.south) coordinate[pos=0.3] (p-Ren);
  \path (nRS.north west) -- (nRS.north) coordinate[pos=0.3] (p-RS);
  \draw [->, arrow] (nRenamer.south) -- (nReorder.north);
  %\draw [->, arrow] (p-Ren) -- node [left] {4--6 \textnormal\microops} +(0,-0.6) -- (p-RS);
  \draw [->, arrow] (p-Ren)  -- (p-RS);
  %\draw [<->, arrow] (nReorder.south) -- (nRS.north);

  \draw [->, arrow] (nReorder.east) -- node [above] {\textnormal retire} +(1.4,0) -- +(1.5,0);
  
  \draw [->, arrow, font=\fontsize{6}{7.2}\sffamily] (nPort0.south) -- node [right] {\textnormal\microop} +(-0,-0.5) -- (nPort0FU.east);
  \draw [->, arrow, font=\fontsize{6}{7.2}\sffamily] (nPort1.south) -- node [right] {\textnormal\microop} +(-0,-0.5) -- (nPort1FU.east);
  \draw [->, arrow, font=\fontsize{6}{7.2}\sffamily] (nPort2.south) -- node [right] {\textnormal\microop} +(-0,-0.5) -- (nPort2FU.east);
  \draw [->, arrow, font=\fontsize{6}{7.2}\sffamily] (nPort3.south) -- node [right] {\textnormal\microop} +(-0,-0.5) -- (nPort3FU.east);
  \draw [->, arrow, font=\fontsize{6}{7.2}\sffamily] (nPort4.south) -- node [right] {\textnormal\microop} +(-0,-0.5) -- (nPort4FU.east);
  \draw [->, arrow, font=\fontsize{6}{7.2}\sffamily] (nPort5.south) -- node [right] {\textnormal\microop} +(-0,-0.5) -- (nPort5FU.east);

  \draw [<->, arrow] (nPort2FU.west) -- (nPort2FU.west |- nL1D.north);
  \draw [<->, arrow] (nPort3FU.west) -- (nPort3FU.west |- nL1D.north);
  \draw [->, arrow] (nPort4FU.west) -- (nPort4FU.west |- nL1D.north);

  \draw [<->, arrow] (nL1D.south) -- (nL2.north);
  \draw [->, arrow] (nL2.east) -- +(+0.8,-0.0) |- (nIC.east);
   
  \background{nIC}{nIC}{nL2}{nIDQ}{-0.5}{{rgb,255:red,255; green,246; blue,213}}{Front End}
  \background{nRS}{nRenamer}{nL2}{ee_br}{-0.5}{{rgb,255:red,213; green,255; blue,230}}{Execution Engine (Back End)}
  \background{nL1D}{nL1D}{nL1D}{nL2}{-0.4}{{rgb,255:red,252; green,222; blue,212}}{Memory}
\end{tikzpicture}
\caption{Pipeline of Intel Core CPUs}
\labfig{pipeline}
\end{figure}
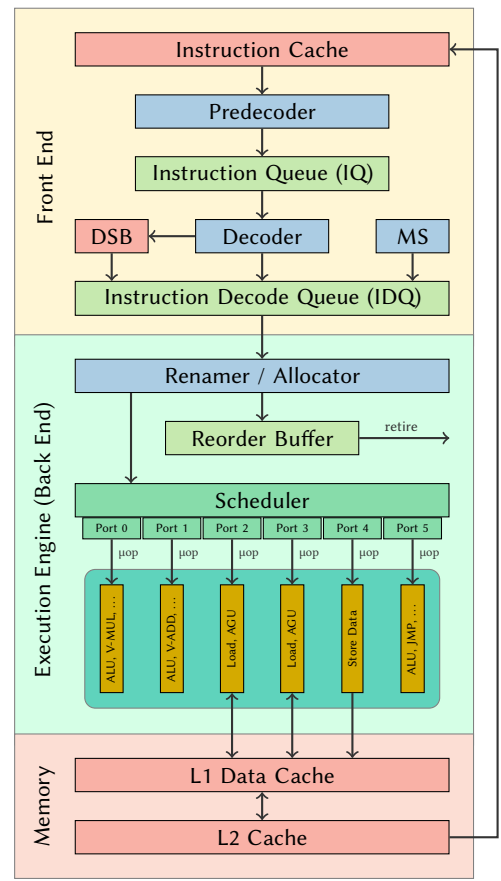

\begin{newInfo}
According to our experiments, the predecoder can predecode at most five instructions per cycle.
If there are, e.g., six instructions in a 16-byte block, in the first cycle, five instructions would be predecoded, and in the next cycle, only one instruction would be predecoded.
Several sources incorrectly claim that the predecoding limit is six instructions per cycle \cite{Schaik19, wikichipSkylake, realworldtechHaswell, ren21}; the source for this might be a section in Intel's optimization manual~\cite{intelOptManual20} that mentions such a limit; however, this section only applies to old Intel Core 2 CPUs, which did not have a \microop cache, and thus decoding was a more significant bottleneck.
\end{newInfo}

A special case are instructions with a so called length-changing prefix (LCP). 
For such instructions, the predecoder has to use a slower length-decoding algorithm.
This results in a penalty of three cycles for each such instruction.

\begin{newInfo}
For building an accurate simulator, it is important to know how instructions that cross a 16-byte boundary are predecoded; however, this is undocumented.
Our experiments show that instructions are predecoded with the 16-byte block in which they end; they count toward the limit of five instructions in the corresponding cycle.
We found that there is a one cycle penalty if five instructions were predecoded in the current cycle, the next instruction crosses a 16-byte boundary, but the primary opcode is contained in the current 16-byte block; if there are only prefixes or escape opcodes of the next instruction in the current block, there is no penalty.
\end{newInfo}

The decoding unit fetches up to four instructions per cycle from the~IQ.
It decodes the instructions into a sequence of \microops and sends them to the \emph{Instruction Decode Queue} (IDQ). %, sometimes also called \emph{\microop queue}).

The decoding unit consists of one complex decoder and three simple decoders.
The simple decoders can only decode instructions with a single \microop.
The complex decoder always decodes the first of the fetched instructions, and it can generate up to four \microops. 
Several sources (e.g., \cite{Schaik19, wikichipSkylake, anandtechSkylake, deng22}) incorrectly claim that with the Skylake microarchitecture, the number of simple decoders was increased from three to four; this might be based on a misinterpretation of the fact that with Skylake, the number of \microops that can be delivered from the decoding unit to the IDQ was increased from four to five.

Instructions with more than four \microops are (at least partially) handled by the Microcode Sequencer (MS).
\begin{newInfo}
We discovered that there are instructions that have only one \microop but that can only be handled by the complex decoder.
\end{newInfo}

\begin{parameter}
The sizes of the IQ and the IDQ, whether a macro-fusible instruction can be decoded on the last decoder or when the instruction queue is empty, whether \verb|pop| instructions require the complex decoder if registers \verb|rsp| or \verb|r12| are used, and whether any other instructions can be decoded in the same cycle are parameters of our model.
\end{parameter}

Decoded \microops are also stored in the \emph{Decoded Stream Buffer} (DSB, also: µOP cache), subject to certain conditions.
This can allow for a higher throughput of loops for which decoding is the bottleneck.

A cache line in the DSB of a pre-Ice Lake CPUs can store at most six \microops that need to be contained in the same 32-byte aligned code block.
There can be at most three cache lines that contain \microops from a specific 32-byte block.
If a 32-byte block contains more than three cache lines, no \microop of this block will be stored in the DSB.
There are a number of other restrictions; e.g., some \microops require two slots in a cache line.
These restrictions are not described in the official manuals, but they have been reverse engineered by Agner Fog~\cite{fog21uArch}.

\begin{newInfo}
We have discovered that on Skylake and Cascade Lake CPUs, \microops from a specific 32-byte block are only served from the DSB if both 32-byte blocks of the corresponding 64-byte instruction cache line fulfill the restrictions described in the previous paragraph.

Starting with the Ice Lake microarchitecture, the DSB operates on 64-byte blocks.
There can be at most six cache lines (with up to six \microops each) from a specific 64-byte block.
\end{newInfo}

As a workaround for the ``Jump Conditional Code'' (JCC) erratum, Skylake-based CPUs with a recent microcode cannot cache blocks that contain a jump instruction that crosses or ends on a 32-byte boundary~\cite{jccErratum}.													 

\begin{newInfo}
According to~\cite{fog21uArch}, the ``pipeline switches frequently between taking \microops from the decoders and from the \microop cache''. 
We have found out that a switch from the decoders to the DSB can only take place after a branch instruction.
Thus, for loops that contain only one branch instruction at the end, a switch to the DSB can only take place at the start of a loop iteration.

We discovered that when switching from the decoders to the MS and back, there are two stall cycles (in total).
Switching from the DSB to the MS and back incurs two stall cycles on Skylake and its successors, and four stall cycles on earlier microarchitectures; this contradicts~\cite{fog21uArch}, which claims that ``each switch may cost one clock cycle'' on Sandy Bridge and Ivy Bridge.
\end{newInfo}

\begin{parameter}
The block size of the DSB, the maximum number of \microops that the DSB can deliver per cycle, whether a 32 byte block can only be in the DSB if the other 32 byte block in the same 64 byte block is also cacheable, the number of stall cycles when switching from DSB to MS, and whether a branch instruction can be the last instruction in a block that is cached by the DSB are parameters of our model. % (see Jcc erratum).
\end{parameter}

%\negvspacesmall

The \emph{Loop Stream Detector} (LSD) detects loops whose \microops fit entirely into the IDQ.
In this case, it locks the \microops in the IDQ, and streams them continuously without requiring the DSB or the decoders.
The first \microop of a loop iteration cannot be streamed in the same cycle as the last \microop of the previous iteration.
As this could be a significant bottleneck for small loops, the LSD can automatically unroll the loop.
While this unrolling is briefly mentioned in the manual, no details are provided.

\begin{newInfo}
We have reverse engineered how often the LSD unroll loops of different sizes on different microarchitectures.
\end{newInfo}

%\negvspacesmall

\begin{parameter}
Whether the LSD is enabled (on Skylake-based CPUs, it was disabled with a microcode update due to the SKL150 erratum~\cite{sklSpecUpdate}), and how the code is unrolled by the LSD, are parameters of our model.
\end{parameter}

%\negvspacesmall

\subsubsection{Renamer / Allocator}\labsec{renamer}
The \emph{renamer} (also called \emph{Resource Allocation Table} (RAT)) maps architectural registers to physical registers.
It also allocates resources for loads and stores, and it assigns execution ports to the \microops.

The renamer fetches \microops from the IDQ.
It stores all \microops in the reorder buffer, and it issues them to the scheduler (see \refsec{scheduler}).
We call the maximum number of \microops that the renamer can handle per cycle the \emph{issue width}.

All \microops remain in the reorder buffer until they are ready to retire.
A \microop is ready to retire if it has finished execution and all older \microops (in program order) are ready to retire.

The renamer can directly execute certain classes of \microops like register moves (see \refsec{moveElimination}), NOPs, or zero (one) idioms; such \microops are sent to the reorder buffer but not to the scheduler.
Zero (one) idioms are instructions that always set the target register to 0 (1), independently of the values in the source registers. 
An example is an \verb|XOR| of a register with itself.

\begin{parameter}
The size of the reorder buffer, the issue width, the number of instructions that can be retired per cycle, and whether the high 8-bit register are renamed separately from the low 8-bit registers are parameters of our model.
\end{parameter}

%\negvspacesmall

\begin{newInfo}
While it is known from prior work~\cite{Abel19} which ports a \microop may be assigned to, it has been unknown how the renamer chooses the actual ports at runtime.

We have reverse engineered the port assignment algorithm.
In the following, we describe our findings for CPUs with eight ports; such CPUs are currently most widely used.
These CPUs can issue up to four \microops per cycle.

In the following, we will call the position of a \microop within a cycle an \emph{issue slot}; e.g., the oldest instruction issued in a cycle would occupy issue slot 0.

The port that a \microop is assigned depends on its issue slot and on the ports assigned to \microops that have not been executed and were issued in a previous cycle.
In the following, we will only consider \microops that can use more than one port.

For a given \microop $m$, let $P_{min}$ be the port to which the fewest non-executed \microops have been assigned to from among the ports that $m$ can use. 
Let $P_{min'}$ be the port with the second smallest usage so far.
If there is a tie among the ports with the smallest (or second smallest, respectively) usage, let $P_{min}$ (or $P_{min'}$) be the port with the highest port number from among these ports (the reason for this choice is probably that ports with higher numbers are connected to fewer functional units).
If the difference in the usage between $P_{min}$ and $P_{min'}$ is greater or equal to $3$, we set $P_{min'}$ to $P_{min}$.
The \microops in issue slots 0 and 2 are assigned to port $P_{min}$. 
The \microops in issue slots 1 and 3 are assigned to port $P_{min'}$. 

A special case are \microops that can use port~2 and port~3.
These ports are used by \microops that handle memory accesses, and both ports are connected to the same types of functional units.
For such \microops, the port assignment algorithm alternates between port~2 and port~3.
\end{newInfo}

%\negvspacesmall

\subsubsection{Scheduler}\labsec{scheduler}
The scheduler (also called the reservation station) keeps track of the dependencies of the \microops.
Once all operands of a \microop are ready, the scheduler dispatches it to its assigned port.
%\todo{do ports have queues associated with them? how else does the scheduler deal with multiple \microops that are ready in the same cycle and assigned to the same port?}

Each port is connected to a set of different functional units, such as an ALU, an address-generation unit (AGU), or a unit for vector multiplications.
Each port can accept at most one \microop in every cycle. 
However, as most functional units are fully pipelined, a port can typically accept a new \microop in every cycle, even though the corresponding functional unit might not have finished executing a previous \microop.%\looseness=-1
\begin{parameter}
The number of entries of the scheduler, and whether memory loads are 1 cycle faster if a non-indexed addressing mode is used and the base register was written by a move (from memory to register) or a pop instruction are parameters of our model.
\end{parameter}

\subsubsection{Move Elimination}\labsec{moveElimination}
Starting with the Ivy Bridge microarchitecture, certain register-to-register move instructions can be executed by the renamer (see~\refsec{renamer}).

However, this \emph{move elimination} is not always successful.
Intel's manual~\cite{intelOptManual20} mentions ``internal resource constraints'' that may prevent eliminations, and provides an example in which only 50\% of the moves could be eliminated, but it does not describe these internal resources in more detail.

The relevant Intel CPUs have performance counters that count the number of eliminated and non-eliminated move instructions.
We have developed microbenchmarks that use these counters to analyze when move elimination is successful.

\begin{newInfo}
The following model agrees with our observations.
The processor keeps track of the physical registers that are used by more than one architectural register.
We say that each such physical register occupies one \emph{elimination slot}.
An elimination slot is released again after the corresponding registers have been overwritten.
The number of move instructions that can be eliminated in a cycle depends both on the number of available elimination slots, and on the number of successful eliminations in the previous cycle.
%it's somewhat implicit that the number of elimination slots is limited

We have discovered that on Tiger Lake and Ice Lake CPUs with a recent microcode, move elimination for general-purpose registers is disabled.
On Ice Lake CPUs with an older microcode, move elimination is enabled.
This is probably due to the ICL065 erratum~\cite{iclSpecUpdate}.
\end{newInfo}

\begin{parameter}
The number of move elimination slots for general-purpose and SIMD registers, the pipeline length of the move elimination mechanism, whether all aliases to a general-purpose register need to be overwritten before an elimination slot is released, and whether a \verb|movzx| instruction can be eliminated if the second register has the same encoding as a high 8-bit register are parameters of our model.
\end{parameter}

\subsubsection{Macro Fusion}
The Instruction Queue (IQ) can merge specific pairs of instructions; such ``macro-fused'' instruction pairs are treated as single instructions in the rest of the pipeline.
The first instruction of such a pair is always an arithmetic or logic instruction that updates the status flags, and the second instruction is a conditional jump instruction.

%\subsubsection{Stack Engine}

%\subsubsection{Register-Merge \microops}

\subsubsection{Micro Fusion}
Micro fusion is an optimization in which two \microops of the same instruction are fused together in the decoding stage and treated as one \microop in the early parts of the pipeline; they are split into two \microops before execution in the back end.
All decoders (including the simple decoders) can emit micro-fused \microops.
Only specific types of \microops can be fused; one of the \microops must be a load or store \microop.

There are two possible locations in the pipeline where micro-fused \microops may be split into their components.
In most cases, micro-fused \microops are split when they enter the scheduler; however, they take only one slot in the reorder buffer.
In some cases, micro-fused \microops from instructions that use an indexed addressing mode are split already by the renamer; Intel's optimization manual refers to this as ``unlamination''.
Unlaminated \microops require two slots in the reorder buffer.

\begin{newInfo}
We have found out that if the number of \microops after unlamination exceeds the issue width, the renamer issues both \microops that were part of the fused \microop in the next cycle. 
\end{newInfo}

\subsection{Properties of Individual Instructions}\labsec{instructionProperties}
While the pipeline components are relatively similar in different microarchitectures, how instructions are executed, and which instructions are supported can differ significantly.
Recent work~\cite{Abel19} has proposed techniques to automatically determine latency, throughput, and port usage data of individual instructions.
While such data is necessary for constructing an accurate throughput predictor, it is not sufficient.

We have therefore extended the techniques from~\cite{Abel19} so that they are also able to automatically determine the following properties of x86 instructions: 

\begin{newInfo}
  \begin{itemize}
\item How many \microops of an instruction are micro fused, and whether they are unlaminated by the renamer.
\item How many \microops of an instruction are delivered from the decoders (or the DSB) and how many from the MS.
\item Whether an instruction requires the complex decoder. %We discovered that there are instructions that have only one \microop but that can only be handled by the complex decoder.%, which was previously unknown.
\item For each instruction that requires the complex decoder, we determine the number of instructions that can be handled by simple decoders in the same cycle.
\item The pairs of instructions that can be macro fused (these pairs are microarchitecture-specific and undocumented for post-Haswell processors).
\end{itemize}
\end{newInfo}
We have upstreamed our extensions to the open-source repository of~\cite{Abel19}, and the results are available at \href{https://www.uops.info/}{uops.info}.

\begin{figure*}
\includegraphics[width=.95\textwidth]{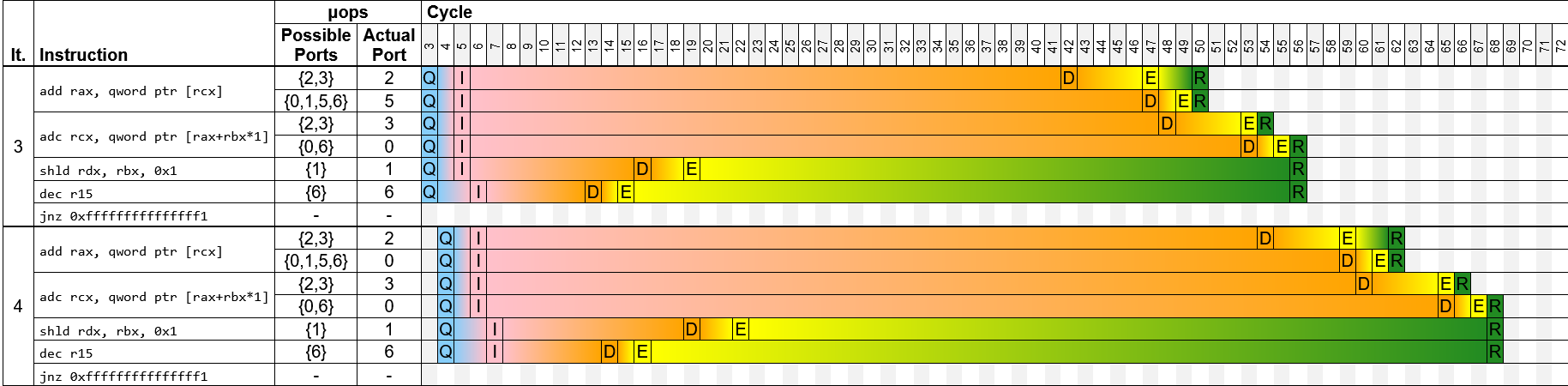}
\caption{Screenshot of uiCA's timeline view}
\labfig{timeline}
\end{figure*}

\subsection{Basic-Block Throughput Predictor}\labsec{TPPredictor}
Based on the model described in the previous paragraphs, we have implemented \uiCA, a tool that simulates the execution of basic blocks on Intel Core microarchitectures.
The tools provides throughput predictions, as well as further insights into how the code is executed, which may be useful for performance engineers. 
Specifically, it can generate a table that contains the actual port usage for each instruction, and it can output a timeline that shows what happens in each cycle; an excerpt from such a timeline is shown in \reffig{timeline}.\looseness=-1

Throughput predictions for a basic block are obtained as follows.
The tool simulates the repeated execution of the basic block for at least 500 cycles and until at least 10 iterations have been completed.
Let $n$ be the number of completed iterations.
Let $t$ be the cycle in which the last instruction of the $n$-th iteration was retired, and~$t'$ be the cycle in which the last instruction of the $\frac{n}{2}$-th iteration was retired.
The tool then predicts $\frac{2\cdot(t-t')}{n}$ as the throughput.
This approach is similar to an approach that was proposed in~\cite{Abel20} for performing measurements.
It is based on the assumption that after $\frac{n}{2}$ iterations, a steady state has been reached.

\section{Benchmarks and Measurements}\labsec{benchmarks}
To evaluate and compare our predictor to previous approaches, we need a set of suitable benchmarks. 
Chen et al.~\cite{Chen19} proposed the BHive benchmark suite, which is designed specifically to evaluate basic-block throughput predictors on x86 systems.
The BHive suite contains more than 300,000 basic blocks that were extracted from applications from different domains, including numerical computation, databases, compilers, machine learning, and cryptography.

In addition to the BHive suite, Chen et al.\ in the same paper also propose a profiling tool to measure the throughput of such basic blocks using hardware performance counters.

While their benchmark suite and measurement framework are in principle suitable for evaluating the work presented in our paper, we discovered a number of issues with their approach that can lead to incorrect or misleading results. 
Thus, in this section, we describe how to overcome these issues.

%\todo{brief outline of section}

\subsection{In-Scope Benchmarks}
The goal of the BHive benchmark suite is to consist of basic blocks whose execution conforms to the common modeling assumptions of throughput predictors discussed in \refsec{assumptions}.

The benchmark suite was originally generated as follows: Chen et al.~\cite{Chen19} first extracted a large number of basic blocks from different applications.
Then, they filtered out benchmarks that are not in scope because they violate common modeling assumptions. 
We have identified several additional benchmarks that violate such modeling assumptions, and we have therefore extended Chen et al.'s filtering approach accordingly.

\paragraph{TLB Misses}
A common modeling assumption of existing throughput predictors is that all memory accesses lead to cache and TLB hits. 
Chen et al. only filter out blocks with cache misses. 
We additionally filter out blocks with TLB misses, as such blocks are out-of-scope for the BHive benchmark suite.
This was confirmed by the authors of the BHive suite.

\paragraph{Unbalanced x87 Operations}
The BHive suite contains several basic blocks that contain an unbalanced number of push or pop operations to the x87 floating-point stack.
Correct programs would not execute such blocks repeatedly in isolation, as this leads to stack underflows or overflows, which result in penalties of hundreds of cycles.
It is worth noting that the basic blocks in the BHive benchmark suite are not necessarily executed in loops in the applications from which they were extracted, and thus underflows or overflows would not occur in their original contexts.

A common modeling assumption is that no underflows or overflows would occur. 
We therefore filtered out the corresponding benchmarks, as they are not in scope for the benchmark suite.

\paragraph{Unsupported Instructions}
The \verb|TZCNT| instruction was introduced with the Haswell microarchitecture.
It has the same encoding as the \verb|BSF| instruction with a \verb|REP| prefix.
This prefix is undefined for the \verb|BSF| instruction; however, older CPUs do not generate an exception in this case, but simply ignore the prefix.

We removed the \verb|TZCNT| instruction from the benchmarks for older microarchitectures, as it is not meaningful to evaluate throughput predictions on unsupported instructions.

\paragraph{Benchmarks with Input-Dependent Timing}
As discussed in \refsec{assumptions}, throughput predictors make assumptions on the inputs that are used for basic blocks that have input-dependent timing.
However, the inputs that the BHive profiler uses do, in general, not conform to these assumptions.
For example, the BHive profiler initializes all register with the same value, which often leads to memory aliasing.
It should be noted that in the context from which the basic blocks were originally extracted, it would rarely be the case that all registers have the same value.
Moreover, division instructions that use these values are typically fast, which conflicts with the pessimistic assumptions of most throughput predictors for such cases.\looseness=-1

Ideally, one should choose initial values such that the execution conforms to the modeling assumptions.
However, developing an approach to do so automatically is beyond the scope of this paper; doing it manually is infeasible due to the large number of benchmarks.\looseness=-1

Using measurements with the currently used input values to compare throughput predictors is not very meaningful, as it could give an unfair advantage to the tools that were trained on measurements with the same input values. 
It would also give an unfair advantage to our predictor, as we know the input values that the BHive profiler uses, and thus we could optimize our predictions for these inputs.%\looseness=-1

We therefore filter out benchmarks that use the \verb|DIV|, \verb|SQRT|, and \verb|CPUID| instructions, as well as benchmarks for which it may depend on the inputs whether there are read-after-write dependencies. %that seems quite drastic
However, we do not filter out benchmarks that use the same address registers for reading and writing to memory and that do not modify this register, as these benchmarks always have a read-after-write dependency, independently of the initial values in the address registers.
Furthermore, for the microarchitectures to which it applies, we also filter out benchmarks for which it depends on the inputs whether there is a bank conflict, or whether stores can be paired.%\looseness=-1

%\paragraph{State of Registers}
%Last write for vector instructions/clean upper state
%last write for high8 registers

\subsection{Loop-Based Benchmarks}
The benchmarks in the BHive suite do not end in branch instructions.
The BHive profiler measures their throughput according to the \DefTPU definition (see \refsec{definitions}).
In~\cite{Chen19, mendis19a}, these measurements are used to compare the predictions of Ithemal (which was trained on benchmarks that were evaluated with the same profiler) to the predictions of IACA, OSACA, and llvm-mca (which are based on the \DefTPL definition).
As the predictions of Ithemal are closer to the measurements than the predictions of the other tools, they conclude that Ithemal ``outperforms'' the other tools.
We don't think this conclusion is valid because the measurements are based on a different definition of throughput than the predictions of the other tools.\looseness=-1

In order to enable a more meaningful comparison with previous tools, we have created a variant of the BHive benchmark suite in which the benchmarks end in a branch instruction, so that they are applicable to the \DefTPL definition. In the following, we will call this benchmark suite \bhivel, and the original benchmark suite \bhiveu.

We have generated the benchmarks in \bhivel from those in \bhiveu as follows. 
Let $B$ be a benchmark in \bhiveu, and let $R_x$ be a general-purpose register that is not used by $B$.
We then add to \bhivel an extended benchmark of the form
\begin{center}
\texttt{loop: $B$; DEC $R_x$; JNZ loop}
\end{center}
Here, $R_x$ is used as the loop counter.
For a small number of benchmarks in \bhiveu, we could not find such a register $R_x$, as these benchmarks already use all general-purpose registers. We omitted these benchmarks from \bhivel.

Several of the benchmarks in \bhiveu are very small; some of them consist of only a single instruction.
For such benchmarks, the execution time of the extended benchmark may be dominated by the loop overhead, which limits the throughput to one iteration per cycle in the best case.
Therefore, for benchmarks $B$ with fewer than five instructions, we added an additional variant in which $B$ is unrolled until there are at least five instructions (which corresponds to the maximum issue width of the CPUs that we consider) in the body of the loop.

%We have extended the BHive profiler to support evaluating the benchmarks in \bhivel.

\subsection{Performing Accurate Measurements}\labsec{measurements}
For a meaningful comparison of measurements to predictions, it is important that the measurements are performed in an accurate way and in a well-defined setting.

Based on our experience, measurements using hardware performance counters can be almost cycle-accurate if some precautions are taken.
Unfortunately, the BHive profiler tool does not always achieve this accuracy.
Skylake CPUs, for example, can execute at most two instructions of the same kind with memory operands per cycle. Thus, any measurement with a throughput value smaller than 0.5 cycles per iteration for benchmarks with memory instructions is obviously not accurate.
The published measurements file, which was used for the evaluation in~\cite{Chen19}, contains almost 20,000 such cases. More than 2,200 of them even report a throughput value of less than 0.45 cycles per iteration (i.e., the measurements are more than 10\% off from the correct value).

One of the reasons for these inaccuracies is that for small basic blocks, the BHive profiler does not use a large enough number of repetitions to actually reach a steady state in all cases.
We use the following approach instead.
Let~$n$ be the number of instructions in a benchmark, and let $r := \left\lceil\frac{500}{n}\right\rceil$.
For \bhiveu, we determine the throughput as the difference between the measured execution times for $r$ and $2\cdot r$ many repetitions, divided by $r$.
This leads to a significantly higher repetition count compared to the original BHive profiler for most benchmarks, but it is still small enough so that the code fits in the instruction cache.
For \bhivel, we use the difference between $10,000$ and $20,000$ iterations, divided by $10,000$.
We perform all measurements with all but one core disabled to prevent disturbances from other processes, and we repeat all throughput measurements 100 times.
We then remove the top and bottom 20\% of the measured values, which might be outliers due to, e.g., interrupts.
Similar to Chen et al., we filter out benchmarks for which the measurements were not stable.
Specifically, we filter out benchmarks if the minimum and the maximum of the remaining throughput values differs by more than 0.02 cycles.
Otherwise, we use the median of the measurements as the throughput.

Another reason for the inaccuracies is that the BHive profiler executes multiple branch instructions on the critical path (i.e., while performance counting is active). 
This makes the execution time dependent on the state of the branch predictor, which, in general, leads to unpredictable measurements.
Furthermore, the \verb|CPUID| instruction is used for serialization, which is relatively expensive and has an input-dependent throughput.
We instead use the \verb|LFENCE| instruction for serialization, as recommended in recent work~\cite{McCalpin18, Abel20}, and we removed all branches from the critical path (except, of course, for branches that the benchmarks themselves contain).

The measured throughput can also depend on the initial state of the microarchitecture. 
For a meaningful comparison of measurements to predictions of previous tools, it is important to perform the measurements under conditions that do not contradict assumptions made by these tools.
Specifically, we perform all benchmark runs under the following initial conditions.
We make sure that all move elimination resources are available by overwriting all registers, we drain all front-end buffers by executing a long enough sequence of 15-byte NOP instructions, and we align the first instruction of the benchmark to a 64-byte cache line.

\begin{comment}
\subsection{Size of Improved Benchmark Set}
After excluding the benchmarks described in the previous sections, \bhiveu contains between $135,915$ (for Ivy Bridge) and $193,777$ (for Tiger Lake) benchmarks, and \bhivel contains between $330,503$ (for Rocket Lake) and $352,118$ (for Cascade Lake) benchmarks.
This is still a very large number of benchmarks, which allows for a meaningful comparison of different throughput prediction tools.
\todo{possibly contrast \bhiveu numbers with original values?}
\end{comment}

%Incorrect measurements (e.g. unroll count too small, example from Github issue, etc.)

%Branch instructions on critical path

%Non-issues:

%warm-up for AVX (ToDo: SSE/AVX transition penalties?)

\section{Experimental Evaluation}\labsec{evaluation}

\subsection{Comparison with Other Tools}

In this section we compare our tool, \uiCA, with several previous tools on all major Intel Core microarchitecture generations that were released in the last ten years, from Sandy Bridge (released in 2011) to Rocket Lake (released in 2021).

We use IACA~\cite{iacaGuide} in versions 2.3 and 3.0; version 3.0 does not support the older microarchitectures, and we noticed that IACA 2.3 tends to provide more accurate predictions.
For llvm-mca, we use version 10.0.0; for the microarchitectures that are supported by DiffTune, we additionally also evaluate llvm-mca in version 8.0.1, as DiffTune is based on this version.
We use DiffTune~\cite{Renda20} at commit 9992f69 with the models from the paper, which are provided at\footnote{https://github.com/ithemal/DiffTune/issues/1}.
We use Ithemal at commit 47a5734 with the retrained models from the BHive paper. %\footnote{https://github.com/ithemal/Ithemal-models/tree/master/bhive}.
For CQA, we use version 2.13.2.
We use OSACA at commit 63563ec; we do not use the latest released version of OSACA, as we found several bugs in this version that we reported and that were since fixed by the authors in the version we use.
In cases in which a tool crashes or does not return a result within a timeout of one hour, we consider the throughput prediction to be $0$.%\looseness=-1

To compare the different tools, we use the same metrics that were used in ~\cite{Chen19, Renda20}:
\begin{itemize}
\item The mean absolute percentage error (MAPE) of the predictions relative to the measurements, which is defined as follows.
Let $B$ be a set of pairs $(m, p)$ such that $m$ is the measured throughput of a benchmark, and $p$ is the predicted throughput.
Then $$MAPE(B) = \frac{1}{\vert B \vert} \cdot \sum_{(m,p) \in B} \frac{\vert m - p \vert}{m}$$
\item Kendall's tau coefficient~\cite{Kendall38}, which is a measure for how well the pairwise ordering is preserved.
As argued by Chen et al.~\cite{Chen19}, Kendall's tau coefficient can be more useful than the MAPE for example for compiler optimizations, where the relative ordering of different candidate programs is more important than their absolute execution times.
\end{itemize}

As a baseline, we use the following two simple analytical throughput prediction models corresponding to the two throughput definitions discussed in \refsec{definitions}.
Let $n$ be the number of instructions in a specific benchmark, and let $m_r$ and $m_w$ be the number of memory read and write accesses of the benchmark.
Furthermore, let $i$ be the issue width (see~\refsec{renamer}) of the corresponding microarchitecture, and let $w$ be the number of memory write operations that can be performed per cycle.

For the basic blocks in \bhiveu, we use \[TP_{baseline, U} = max\left(\frac{n}{4}, \frac{m_r}{2}, \frac{m_w}{w}\right)\] as the baseline (this is a generalization to other microarchitectures of the baseline that we discussed in the introduction). %\todo{what share of benchmarks is ``bottlenecked'' by each of the five components? this would be quite interesting to understand the set of benchmarks}
This value constitutes a lower bound on the execution time of basic blocks without branch instructions, as at most four instructions can be decoded per cycle, and at most two memory read operations can be performed per cycle.

For the benchmarks in \bhivel, we use \[TP_{baseline, L} = max\left(1, \frac{n-1}{i}, \frac{m_r}{2}, \frac{m_w}{w}\right)\] as the baseline.
We do not include a term for the decoding limit here, as the \microops for the benchmarks in \bhivel are often delivered from the DSB or the LSD.
Instead, $\frac{n-1}{i}$ corresponds to the issue limit; we use $n-1$ instead of $n$, as the last two instructions of such basic blocks are often macro fused. 
We do not include a term corresponding to the issue limit for $TP_{baseline, U}$, as $i \geq 4$ for all considered microarchitectures.
We use $1$ as an additional lower bound, as the benchmarks in \bhivel cannot run faster than one iteration per cycle due to the read-after-write dependency of the decrement operation.
Note that the only microarchitecture-specific variables in these formulas are $i$ and $w$; all other variables only depend on the benchmark and are independent of the microarchitecture.

\reftab{compBHive} shows the results of our evaluation.
For completeness and consistency with the evaluations in~\cite{Chen19, Renda20}, we also evaluated the tools that are based on the \DefTPL throughput notion on \bhiveu (except for CQA, which can only analyze code that ends in a branch instruction); the corresponding entries are printed in gray.
For DiffTune, it is not clear which throughput notion is the more meaningful one: DiffTune is essentially llvm-mca (which is based on \DefTPL), but it was trained on measurements obtained according to \DefTPU.\looseness=-1

%\reffig{heatmaps} shows heatmaps for \bhivel on Skylake that relate the predicted throughput of different tools to the measured throughput.

In most cases, the accuracy on \bhivel is higher than on \bhiveu; the exception here are Ithemal and DiffTune, which were trained on measurements that were obtained by unrolling.
This shows that, for a meaningful comparison, it is important that measurements and predictions are based on the same definition of throughput (as discussed in \refsec{definitions}).

\begin{table}
\caption{Microarchitectures used for the evaluation}
  \labeltab{uarch}
  \begin{center}
  \begin{tabular}{lccl}
  \toprule
  \textbf{{$\mu$}Arch} & \textbf{Abbr.} & \textbf{Released} & \textbf{CPU} \\
  \midrule
  Rocket Lake & RKL & 2021 & Intel Core~i9-11900 \\
  Tiger Lake & TGL & 2020 & Intel Core~i7-1165G7 \\
  Ice Lake & ICL & 2019 & Intel Core~i5-1035G1 \\
  Cascade Lake & CLX & 2019 & Intel Core i9-10980XE \\
  Skylake & SKL & 2015 & Intel Core~i7-6500U \\
  Broadwell & BDW & 2015 & Intel Core~i5-5200U \\
  Haswell & HSW & 2013 & Intel Xeon~E3-1225~v3 \\
  Ivy Bridge & IVB & 2012 & Intel Core~i5-3470 \\
  Sandy Bridge & SNB & 2011 & Intel Core~i7-2600\\
  \bottomrule
  \end{tabular}
  \end{center}
\end{table}

% https://github.com/andreas-abel/uiCA-eval @3b33eb7

\newcommand{\wrongDef}[1]{\textcolor{gray}{#1}}

\begin{table}
\caption{Comparison of different tools on \bhiveu and \bhivel}
\labeltab{compBHive}
\begin{center}
\resizebox*{!}{.975\textheight}{
\begin{tabular}{llrcrc}
\toprule
& & \multicolumn{2}{c}{\textbf{\bhiveu}} & \multicolumn{2}{c}{\textbf{\bhivel}}\\  \cmidrule(lr){3-4}\cmidrule(lr){5-6}
\textbf{{$\mu$}Arch}  & \textbf{Predictor} & \textbf{MAPE} & \textbf{Kendall} & \textbf{MAPE} & \textbf{Kendall} \\
\midrule
\multirow{2}{*}{RKL}  & \uiCA & 0.49\% & 0.9835 & 0.92\% & 0.9755 \\
                      & Baseline & 15.50\% & 0.7397 & 9.26\% & 0.7808\\
\midrule
\multirow{3}{*}{TGL}  & \uiCA & 0.97\% & 0.9769 & 0.98\% & 0.9731 \\
                      & llvm-mca-10 & \wrongDef{25.74\%} & \wrongDef{0.7049} & 13.80\% & 0.8486\\
                      & Baseline & 17.49\% & 0.7245 & 11.25\% & 0.7413\\
\midrule
\multirow{5}{*}{ICL}  & \uiCA & 1.00\% & 0.9771 & 0.77\% & 0.9759 \\
                      & OSACA & \wrongDef{53.80\%} & \wrongDef{0.3143} & 21.98\% & 0.4698 \\
                      & llvm-mca-10 & \wrongDef{25.38\%} & \wrongDef{0.7030} & 13.64\% & 0.8512 \\
                      & CQA & & & 6.74\% & 0.8835 \\
                      & Baseline & 17.54\% & 0.7230 & 10.84\% & 0.7510\\
\midrule
\multirow{4}{*}{CLX}  & \uiCA & 0.45\% & 0.9713 & 0.65\% & 0.9825 \\
                      & llvm-mca-10 & \wrongDef{23.17\%} & \wrongDef{0.7211} & 13.21\% & 0.8060 \\
                      & OSACA & \wrongDef{20.83\%} & \wrongDef{0.7511} & 11.61\% & 0.8068 \\
                      & Baseline & 15.49\% & 0.7461 & 10.31\% & 0.8021\\
\midrule
\multirow{11}{*}{SKL} & \uiCA & 0.45\% & 0.9798 & 0.38\% & 0.9895 \\
                      & Ithemal & 8.28\% & 0.8172 & \wrongDef{13.66\%} & \wrongDef{0.7582} \\
                      & IACA 3.0 & \wrongDef{13.49\%} & \wrongDef{0.7802} & 14.26\% & 0.8290 \\
                      & IACA 2.3 & \wrongDef{11.85\%} & \wrongDef{0.8071} & 8.42\% & 0.8477 \\
                      & OSACA & \wrongDef{14.95\%} & \wrongDef{0.7639} & 11.25\% & 0.8045 \\
                      & llvm-mca-10 & \wrongDef{15.61\%} & \wrongDef{0.7258} & 12.01\% & 0.8015 \\
                      & llvm-mca-8 & \wrongDef{15.39\%} & \wrongDef{0.7434} & 11.98\% & 0.8021 \\
                      & DiffTune & 24.48\% & 0.6626 & 104.88\% & 0.6426 \\
                      & CQA & & & 7.44\% & 0.8847 \\
                      & \emph{Measured~\cite{Chen19}} & 4.40\% & 0.9113 \\
                      & Baseline & 17.28\% & 0.7228 & 10.03\% & 0.7999\\
\midrule
\multirow{7}{*}{BDW}  & \uiCA & 1.08\% & 0.9805 & 0.61\% & 0.9841 \\ 
                      & IACA 3.0 & \wrongDef{14.69\%} & \wrongDef{0.8012} & 11.47\% & 0.8725 \\
                      & IACA 2.3 & \wrongDef{13.22\%} & \wrongDef{0.8206} & 5.84\% & 0.8928 \\
                      & OSACA & \wrongDef{17.52\%} & \wrongDef{0.7456} & 9.69\% & 0.8365 \\
                      & llvm-mca-10 & \wrongDef{14.23\%} & \wrongDef{0.7793} & 16.71\% & 0.8286 \\
                      & CQA & & & 5.03\% & 0.9213 \\
                      & Baseline & 16.97\% & 0.7572 & 7.44\% & 0.8332\\
\midrule
\multirow{11}{*}{HSW} & \uiCA & 0.76\% & 0.9850 & 0.59\% & 0.9842 \\ 
                      & Ithemal & 7.38\% & 0.8400 & \wrongDef{16.19\%} & \wrongDef{0.7700} \\
                      & IACA 3.0 & \wrongDef{15.04\%} & \wrongDef{0.8080} & 12.00\% & 0.8733 \\
                      & IACA 2.3 & \wrongDef{13.13\%} & \wrongDef{0.8291} & 5.79\% & 0.8925 \\
                      & OSACA & \wrongDef{17.84\%} & \wrongDef{0.7463} & 9.77\% & 0.8307 \\
                      & llvm-mca-10 & \wrongDef{20.29\%} & \wrongDef{0.7835} & 18.97\% & 0.8259 \\
                      & llvm-mca-8 & \wrongDef{21.08\%} & \wrongDef{0.7784} & 19.46\% & 0.8171 \\
                      & DiffTune & 24.80\% & 0.6997 & 138.47\% & 0.6925 \\
                      & CQA & & & 5.08\% & 0.9220 \\
                      & \emph{Measured~\cite{Chen19}} & 2.49\% & 0.9379 \\
                      & Baseline & 17.30\% & 0.7604 & 7.57\% & 0.8314\\
\midrule
\multirow{10}{*}{IVB} & \uiCA & 1.51\% & 0.9608 & 1.12\% & 0.9495 \\ 
                      & Ithemal & 7.08\% & 0.8212 & \wrongDef{12.43\%} & \wrongDef{0.7785} \\
                      & IACA 2.3 & \wrongDef{13.94\%} & \wrongDef{0.7739} & 11.54\% & 0.8271 \\
                      & OSACA & \wrongDef{36.23\%} & \wrongDef{0.4884} & 24.88\% & 0.5846 \\
                      & llvm-mca-10 & \wrongDef{22.79\%} & \wrongDef{0.7656} & 20.76\% & 0.8154 \\
                      & llvm-mca-8 & \wrongDef{22.93\%} & \wrongDef{0.7622} & 20.76\% & 0.8138 \\
                      & DiffTune & 26.21\% & 0.6470 & 82.94\% & 0.7516 \\
                      & CQA & & & 4.05\% & 0.9174 \\
                      & \emph{Measured~\cite{Chen19}} & 3.15\% & 0.9246 \\
                      & Baseline & 18.81\% & 0.7243 & 14.47\% & 0.7670\\
\midrule
\multirow{6}{*}{SNB}  & \uiCA & 1.91\% & 0.9612 & 0.99\% & 0.9649 \\
                      & IACA 2.3 & \wrongDef{11.91\%} & \wrongDef{0.8194} & 9.95\% & 0.8482 \\
                      & OSACA & \wrongDef{36.85\%} & \wrongDef{0.5311} & 24.75\% & 0.5659 \\
                      & llvm-mca-10 & \wrongDef{22.67\%} & \wrongDef{0.8069} & 18.34\% & 0.8455 \\
                      & CQA & & & 4.08\% & 0.9238 \\
                      & Baseline & 20.28\% & 0.7517 & 15.56\% & 0.7577\\
\bottomrule
\end{tabular}}
\end{center}
\end{table}

Our tool, \uiCA, provides the most accurate predictions in all cases. 
In many cases, the MAPE is lower by an order of magnitude or more compared to the best previous tools.
For \uiCA, Kendall's tau coefficient is always higher than for the previous tools; in most cases, it is significantly higher.

On the Cascade Lake, Skylake, Broadwell, and Haswell microarchitectures, the accuracy of OSACA is similar to several other previous tools.
On Ice Lake, Ivy Bridge, and Sandy Bridge, however, the accuracy is significantly below the baseline.
A possible explanation for this is that for these microarchitectures, there is a relatively large number of instructions that are currently not supported by OSACA; in such cases, OSACA simply ignores these instructions.

\begin{table*}
\caption{Influence of the simulation of different microarchitectural components on the prediction accuracy}
\labeltab{evaluiCA}
\begin{tabular}{llrcrc}
\toprule
& & \multicolumn{2}{c}{\textbf{\bhiveu}} & \multicolumn{2}{c}{\textbf{\bhivel}}\\  \cmidrule(lr){3-4}\cmidrule(lr){5-6}
\textbf{{$\mu$}Arch} & \textbf{Predictor} & \textbf{MAPE} & \textbf{Kendall} & \textbf{MAPE} & \textbf{Kendall}\\
\midrule
\multirow{7}{*}{CLX (all benchmarks)} & \uiCA & 0.45\% & 0.9713 & 0.65\% & 0.9825 \\
                      & \uiCA with simple front end & 8.57\% & 0.8602 & 6.23\% & 0.9048\\
                      & \uiCA with simple port assignment & 2.37\% & 0.9280 & 12.20\% & 0.8613\\
                      & \uiCA without micro fusion & 8.77\% & 0.8683 & 3.31\% & 0.9545\\
                      & \uiCA without macro fusion & 0.48\% & 0.9699 & 8.84\% & 0.8863\\
                      & \uiCA without LSD unrolling & 0.45\% & 0.9713 & 6.72\% & 0.9246\\
                      & Baseline & 15.49\% & 0.7461 & 10.31\% & 0.8021\\
\midrule
\multirow{4}{*}{CLX (benchmarks with moves)} & \uiCA & 0.44\% & 0.9801 & 0.45\% & 0.9836 \\
                      & \uiCA without move elimination & 1.71\% & 0.9656 & 1.67\% & 0.9616\\
                      & \uiCA with full move elimination & 0.52\% & 0.9794 & 0.47\% & 0.9846\\
                      & Baseline & 12.99\% & 0.8352 & 9.77\% & 0.8636\\
\bottomrule
\end{tabular}
\end{table*}

Ithemal cannot analyze code that ends in a branch instruction; for the evaluation of Ithemal on \bhivel, we therefore removed the branch instruction, but we kept the instruction that decrements the loop counter in each iteration.
On \bhiveu, Ithemal provides the best predictions among the previous tools; however, the predictions of \uiCA are significantly better.
On \bhivel, several other previous tools provide better predictions than Ithemal; in two cases the MAPE is even below the baseline.
It is likely that retraining Ithemal on measurements obtained with the methodology described in \refsec{benchmarks} would improve its accuracy on \bhivel.
However, we were unable to do so, as the training set is not publicly available.

Unlike in the evaluation in~\cite{Renda20}, DiffTune does not perform better than llvm-mca on our set of benchmarks.
Moreover, DiffTune's accuracy is below the baseline on all supported microarchitectures. %; on the Skylake and Haswell microarchitectures, the MAPE for \bhivel is more than an order of magnitude higher than that of the baseline.
%i think it would be more useful to reiterate why DiffTune is below the baseline now, but was not in their original evaluation

The rows ``Measured~\cite{Chen19}'' in \reftab{compBHive} show the difference of the original timing measurements from~\cite{Chen19} relative to our improved measurements (see \refsec{measurements}).
The MAPE of these measurements compared to our measurements is up to $4.4\%$.
This shows that the measurement methodology can have an important influence on the results.
Moreover, the high level of agreement of our timing measurements with the predictions of our simulator, combined with the fact that our simulator is, unlike the recent machine learning-based approaches, not based on end-to-end timing measurements, gives us confidence that our measurement methodology is more accurate than the previous approach.

\paragraph{Execution Time}
For the benchmarks in \bhiveu on Skylake, \uiCA (which is implemented in Python) requires on average per benchmark around 105 ms, OSACA 1300 ms, IACA 10 ms, llvm-mca 36~ms. For Ithemal it depends: end-to-end it takes around 580 ms; in interactive mode, each additional benchmark requires around 8~ms.\looseness=-1

\subsection{Influence of Different Components on Prediction Accuracy}
We now evaluate how important different components of the model proposed in \refsec{pipelineModel} are for obtaining accurate throughput predictions.
In \reftab{evaluiCA}, we compare different variants of our tool in which parts of the model were replaced by simpler implementations.
We use the Cascade Lake microarchitecture, as on this microarchitecture, both the LSD and move elimination are enabled.

\subsubsection{Simple Front End}
In the first variant, we replace our front-end model with one that is unbounded and can always deliver the maximum number of \microops to the renamer.
This is similar to the models used by many previous tools.
For \bhiveu, this leads to a large increase in the average error. 
For \bhivel, the increase is smaller, which is expected, since for these benchmarks, the \microops are often delivered from the LSD or the DSB, which have a higher bandwidth than the decoders; however, the error is still almost an order of magnitude higher than with our detailed model.%\looseness=-1

\subsubsection{Simple Port Assignment}
For the second variant, we replace the port assignment algorithm that we reverse engineered (see \refsec{renamer}) with one that randomly selects (with uniform probability) a port from among the ports that a \microop can use.
This is similar to the approach described in~\cite{Laukemann18}. %\todo{... followed by OSACA?}
For \bhiveu, this leads to an error that is almost five times as high, and for \bhivel, to an error that is more than 17 times as high.
A main reason for the higher error in the second case is probably that taken branch instructions can only use port~6; with the random port algorithm there is more competition for port~6 from \microops which can also use other ports, and which would more frequently be scheduled on one of these other ports by the actual hardware.
%\todo{would be interesting to evaluate ``optimal'' port assignment}
%\todo{it still appears surprising that some of these simplifications let us drop below the baseline immediately}

\subsubsection{No Micro Fusion}
For the third variant, we assume that \microops from the same instruction cannot be micro fused by the decoders. 
For \bhivel, this leads to an error that is about five times as high.
For \bhiveu, the error is almost 18 times as high; the main reason for this is likely that all instructions that are normally decoded to one micro-fused \microop now require the complex decoder.

\subsubsection{No Macro Fusion}
In the fourth variant, we assume that instructions cannot be macro fused.
For \bhiveu, this makes no difference, as these benchmarks contain no branch instructions.
For \bhivel, on the other hand, the error increases by more than an order of magnitude.

\subsubsection{No LSD Unrolling}
In the next variant, the LSD does not perform unrolling.
Again, this leads to no difference for \bhiveu. %, as these benchmarks do not use the LSD.
For \bhivel, the error is almost an order of magnitude higher.

\subsubsection{Move Elimination}
Finally, we evaluate the influence of the move elimination approach that we reverse engineered in \refsec{moveElimination}. 
For this, we consider only benchmarks that actually contain move instructions, which is the case for more than one third of the benchmarks.
We consider two variants.
In the first variant, no move instructions are eliminated. 
This leads to an average error that is more than three times as high.
In the second variant, all eligible move instructions are eliminated.
For \bhiveu, this leads to an error that is about 15\% higher, and for \bhivel to an error that is about 6\% higher.

\section{Conclusions and Future Work}
Based on a new parametric pipeline model, we have developed an open-source simulator to predict the throughput of basic blocks that is significantly more accurate than the state of the art.
Our experimental evaluation demonstrates that modeling microarchitectural details considered to be rather insignificant in previous work, is in fact crucial for accurate predictions.

Unlike recently proposed machine learning-based techniques, our approach is not based only on end-to-end measurements, but on focused reverse engineering of individual components, which gives a higher confidence that our model is not overfitting to measurement errors.
Training and evaluating predictors on measurements obtained with the same, potentially biased, measurement methodology may result in misleading conclusions, as our results show.

While in this work, we focused on predicting the performance of basic blocks, there is nothing that fundamentally limits our model to basic blocks.
In fact, combining our model with a branch prediction model to support predictions for instruction sequences involving potentially multiple branches would be relatively straightforward.
Similarly, it would be possible to combine it with, e.g., a memory hierarchy simulator to enable predictions that go beyond the typical capabilities of basic-block throughput predictors.
Furthermore, it is conceivable to integrate our model into more comprehensive tools, like full-system simulators, or tools that combine static and dynamic analyses.

\section{Artifacts}
The source code of uiCA, our improved benchmark suite, and the scripts to generate our tables and heatmaps are available on GitHub\footnote{\url{https://github.com/andreas-abel/uiCA}}\footnote{\url{https://github.com/andreas-abel/uiCA-eval}}.
A snapshot of the contents of these repositories at the time of writing is also available on Zenodo~\cite{zenodo}.

Additionally, we provide an interactive online version of our tool at \href{https://uica.uops.info/}{uica.uops.info}.

\begin{acks}
	This project has received funding from the \grantsponsor{1}{European Research Council}{https://erc.europa.eu/}
	under the European Union’s Horizon 2020 research and innovation programme (grant agreement No. \grantnum{1}{101020415}).
\end{acks}

%\todo{analytical throughput predictor}

%\todo{simplicity vs accuracy}

%eval shows that learning based on a potentially biased measurement methodology may give an unfair advantage

%General problem: training based on (good/bad) measurements, will lead to (good/bad) models; evaluating using the same measurement methodology may not be fair (depending on quality of measurements)

%Difference to DiffTune/Ithemal/...: model is not learned end-to-end, which gives higher confidence that model is not overfitting to measurement errors

%Evaluation demonstrates how important it is to model minute details of different components for achieving high accuracy throughput predictions

%Stress again that simulator will be made available open source

%Insights on good measurement practice, e.g. what are the consequences of input- and initial-state-dependent timing for throughput predictors? Should they predict the range of throughputs or, e.g. an average? 

%Future Work:

%AMD

%non-x86

%bottleneck analysis

%visualization

%hyperthreading

%\clearpage

\bibliographystyle{plainurl}
\bibliography{references}

\clearpage

\onecolumn
\appendix 

\section{Heatmaps for Ice Lake}
\vfill
\begin{figure}[H]
\centering
\begin{subfigure}[t]{0.29\textwidth}\resizebox{\textwidth}{!}{\import{heatmaps}{hm_icl_unroll_uiCA.pgf}}\end{subfigure}~
\begin{subfigure}[t]{0.29\textwidth}\resizebox{\textwidth}{!}{\import{heatmaps}{hm_icl_unroll_osaca.pgf}}\end{subfigure}~
\begin{subfigure}[t]{0.29\textwidth}\resizebox{\textwidth}{!}{\import{heatmaps}{hm_icl_unroll_mca.pgf}}\end{subfigure}\\

\begin{subfigure}[t]{0.29\textwidth}\resizebox{\textwidth}{!}{\import{heatmaps}{hm_icl_unroll_baseline.pgf}}\end{subfigure}
\caption{Heatmaps for \bhiveu for basic blocks with a measured throughput of less than 10 cycles/iteration on Ice Lake}
\end{figure}
\vfill

\begin{figure}[H]
\centering
\begin{subfigure}[t]{0.29\textwidth}\resizebox{\textwidth}{!}{\import{heatmaps}{hm_icl_loop_uiCA.pgf}}\end{subfigure}~
\begin{subfigure}[t]{0.29\textwidth}\resizebox{\textwidth}{!}{\import{heatmaps}{hm_icl_loop_osaca.pgf}}\end{subfigure}~
\begin{subfigure}[t]{0.29\textwidth}\resizebox{\textwidth}{!}{\import{heatmaps}{hm_icl_loop_mca.pgf}}\end{subfigure}\\

\begin{subfigure}[t]{0.29\textwidth}\resizebox{\textwidth}{!}{\import{heatmaps}{hm_icl_loop_cqa.pgf}}\end{subfigure}~
\begin{subfigure}[t]{0.29\textwidth}\resizebox{\textwidth}{!}{\import{heatmaps}{hm_icl_loop_baseline.pgf}}\end{subfigure}
\caption{Heatmaps for \bhivel for basic blocks with a measured throughput of less than 10 cycles/iteration on Ice Lake}
\end{figure}

\section{Heatmaps for Skylake}

\vfill
\begin{figure}[H]
\centering
\begin{subfigure}[t]{0.33\textwidth}\resizebox{\textwidth}{!}{\import{heatmaps}{hm_skl_unroll_uiCA.pgf}}\end{subfigure}~
\begin{subfigure}[t]{0.33\textwidth}\resizebox{\textwidth}{!}{\import{heatmaps}{hm_skl_unroll_ith.pgf}}\end{subfigure}~
\begin{subfigure}[t]{0.33\textwidth}\resizebox{\textwidth}{!}{\import{heatmaps}{hm_skl_unroll_iaca3.pgf}}\end{subfigure}\par\bigskip

\begin{subfigure}[t]{0.33\textwidth}\resizebox{\textwidth}{!}{\import{heatmaps}{hm_skl_unroll_iaca23.pgf}}\end{subfigure}~
\begin{subfigure}[t]{0.33\textwidth}\resizebox{\textwidth}{!}{\import{heatmaps}{hm_skl_unroll_osaca.pgf}}\end{subfigure}~
\begin{subfigure}[t]{0.33\textwidth}\resizebox{\textwidth}{!}{\import{heatmaps}{hm_skl_unroll_mca.pgf}}\end{subfigure}\par\bigskip

\begin{subfigure}[t]{0.33\textwidth}\resizebox{\textwidth}{!}{\import{heatmaps}{hm_skl_unroll_difftune.pgf}}\end{subfigure}~
\begin{subfigure}[t]{0.33\textwidth}\resizebox{\textwidth}{!}{\import{heatmaps}{hm_skl_unroll_baseline.pgf}}\end{subfigure}
\caption{Heatmaps for \bhiveu for basic blocks with a measured throughput of less than 10 cycles/iteration on Skylake}
\end{figure}
\vfill

\newpage

\mbox{}
\vfill
\begin{figure}[H]
\centering
\begin{subfigure}[t]{0.33\textwidth}\resizebox{\textwidth}{!}{\import{heatmaps}{hm_skl_loop_uiCA.pgf}}\end{subfigure}~
\begin{subfigure}[t]{0.33\textwidth}\resizebox{\textwidth}{!}{\import{heatmaps}{hm_skl_loop_ith.pgf}}\end{subfigure}~
\begin{subfigure}[t]{0.33\textwidth}\resizebox{\textwidth}{!}{\import{heatmaps}{hm_skl_loop_iaca3.pgf}}\end{subfigure}\par\bigskip

\begin{subfigure}[t]{0.33\textwidth}\resizebox{\textwidth}{!}{\import{heatmaps}{hm_skl_loop_iaca23.pgf}}\end{subfigure}~
\begin{subfigure}[t]{0.33\textwidth}\resizebox{\textwidth}{!}{\import{heatmaps}{hm_skl_loop_osaca.pgf}}\end{subfigure}~
\begin{subfigure}[t]{0.33\textwidth}\resizebox{\textwidth}{!}{\import{heatmaps}{hm_skl_loop_mca.pgf}}\end{subfigure}\par\bigskip

\begin{subfigure}[t]{0.33\textwidth}\resizebox{\textwidth}{!}{\import{heatmaps}{hm_skl_loop_difftune.pgf}}\end{subfigure}~
\begin{subfigure}[t]{0.33\textwidth}\resizebox{\textwidth}{!}{\import{heatmaps}{hm_skl_loop_cqa.pgf}}\end{subfigure}~
\begin{subfigure}[t]{0.33\textwidth}\resizebox{\textwidth}{!}{\import{heatmaps}{hm_skl_loop_baseline.pgf}}\end{subfigure}
\caption{Heatmaps for \bhivel for basic blocks with a measured throughput of less than 10 cycles/iteration on Skylake}
\end{figure}
\vfill

\newpage

\section{Heatmaps for Haswell}

\vfill
\begin{figure}[H]
\centering
\begin{subfigure}[t]{0.33\textwidth}\resizebox{\textwidth}{!}{\import{heatmaps}{hm_hsw_unroll_uiCA.pgf}}\end{subfigure}~
\begin{subfigure}[t]{0.33\textwidth}\resizebox{\textwidth}{!}{\import{heatmaps}{hm_hsw_unroll_ith.pgf}}\end{subfigure}~
\begin{subfigure}[t]{0.33\textwidth}\resizebox{\textwidth}{!}{\import{heatmaps}{hm_hsw_unroll_iaca3.pgf}}\end{subfigure}\par\bigskip

\begin{subfigure}[t]{0.33\textwidth}\resizebox{\textwidth}{!}{\import{heatmaps}{hm_hsw_unroll_iaca23.pgf}}\end{subfigure}~
\begin{subfigure}[t]{0.33\textwidth}\resizebox{\textwidth}{!}{\import{heatmaps}{hm_hsw_unroll_osaca.pgf}}\end{subfigure}~
\begin{subfigure}[t]{0.33\textwidth}\resizebox{\textwidth}{!}{\import{heatmaps}{hm_hsw_unroll_mca.pgf}}\end{subfigure}\par\bigskip

\begin{subfigure}[t]{0.33\textwidth}\resizebox{\textwidth}{!}{\import{heatmaps}{hm_hsw_unroll_difftune.pgf}}\end{subfigure}~
\begin{subfigure}[t]{0.33\textwidth}\resizebox{\textwidth}{!}{\import{heatmaps}{hm_hsw_unroll_baseline.pgf}}\end{subfigure}
\caption{Heatmaps for \bhiveu for basic blocks with a measured throughput of less than 10 cycles/iteration on Haswell}
\end{figure}
\vfill

\newpage

\mbox{}
\vfill
\begin{figure}[H]
\centering
\begin{subfigure}[t]{0.33\textwidth}\resizebox{\textwidth}{!}{\import{heatmaps}{hm_hsw_loop_uiCA.pgf}}\end{subfigure}~
\begin{subfigure}[t]{0.33\textwidth}\resizebox{\textwidth}{!}{\import{heatmaps}{hm_hsw_loop_ith.pgf}}\end{subfigure}~
\begin{subfigure}[t]{0.33\textwidth}\resizebox{\textwidth}{!}{\import{heatmaps}{hm_hsw_loop_iaca3.pgf}}\end{subfigure}\par\bigskip

\begin{subfigure}[t]{0.33\textwidth}\resizebox{\textwidth}{!}{\import{heatmaps}{hm_hsw_loop_iaca23.pgf}}\end{subfigure}~
\begin{subfigure}[t]{0.33\textwidth}\resizebox{\textwidth}{!}{\import{heatmaps}{hm_hsw_loop_osaca.pgf}}\end{subfigure}~
\begin{subfigure}[t]{0.33\textwidth}\resizebox{\textwidth}{!}{\import{heatmaps}{hm_hsw_loop_mca.pgf}}\end{subfigure}\par\bigskip

\begin{subfigure}[t]{0.33\textwidth}\resizebox{\textwidth}{!}{\import{heatmaps}{hm_hsw_loop_difftune.pgf}}\end{subfigure}~
\begin{subfigure}[t]{0.33\textwidth}\resizebox{\textwidth}{!}{\import{heatmaps}{hm_hsw_loop_cqa.pgf}}\end{subfigure}~
\begin{subfigure}[t]{0.33\textwidth}\resizebox{\textwidth}{!}{\import{heatmaps}{hm_hsw_loop_baseline.pgf}}\end{subfigure}
\caption{Heatmaps for \bhivel for basic blocks with a measured throughput of less than 10 cycles/iteration on Haswell}
\end{figure}
\vfill

\newpage

\section{Heatmaps for Ivy Bridge}

\vfill
\begin{figure}[H]
\centering
\begin{subfigure}[t]{0.33\textwidth}\resizebox{\textwidth}{!}{\import{heatmaps}{hm_ivb_unroll_uiCA.pgf}}\end{subfigure}~
\begin{subfigure}[t]{0.33\textwidth}\resizebox{\textwidth}{!}{\import{heatmaps}{hm_ivb_unroll_ith.pgf}}\end{subfigure}~
\begin{subfigure}[t]{0.33\textwidth}\resizebox{\textwidth}{!}{\import{heatmaps}{hm_ivb_unroll_iaca23.pgf}}\end{subfigure}\par\bigskip

\begin{subfigure}[t]{0.33\textwidth}\resizebox{\textwidth}{!}{\import{heatmaps}{hm_ivb_unroll_osaca.pgf}}\end{subfigure}~
\begin{subfigure}[t]{0.33\textwidth}\resizebox{\textwidth}{!}{\import{heatmaps}{hm_ivb_unroll_mca.pgf}}\end{subfigure}~
\begin{subfigure}[t]{0.33\textwidth}\resizebox{\textwidth}{!}{\import{heatmaps}{hm_ivb_unroll_difftune.pgf}}\end{subfigure}\par\bigskip

\begin{subfigure}[t]{0.33\textwidth}\resizebox{\textwidth}{!}{\import{heatmaps}{hm_ivb_unroll_baseline.pgf}}\end{subfigure}
\caption{Heatmaps for \bhiveu for basic blocks with a measured throughput of less than 10 cycles/iteration on Ivy Bridge}
\end{figure}
\vfill

\newpage

\mbox{}
\vfill
\begin{figure}[H]
\centering
\begin{subfigure}[t]{0.33\textwidth}\resizebox{\textwidth}{!}{\import{heatmaps}{hm_ivb_loop_uiCA.pgf}}\end{subfigure}~
\begin{subfigure}[t]{0.33\textwidth}\resizebox{\textwidth}{!}{\import{heatmaps}{hm_ivb_loop_ith.pgf}}\end{subfigure}~
\begin{subfigure}[t]{0.33\textwidth}\resizebox{\textwidth}{!}{\import{heatmaps}{hm_ivb_loop_iaca23.pgf}}\end{subfigure}\par\bigskip

\begin{subfigure}[t]{0.33\textwidth}\resizebox{\textwidth}{!}{\import{heatmaps}{hm_ivb_loop_osaca.pgf}}\end{subfigure}~
\begin{subfigure}[t]{0.33\textwidth}\resizebox{\textwidth}{!}{\import{heatmaps}{hm_ivb_loop_mca.pgf}}\end{subfigure}~
\begin{subfigure}[t]{0.33\textwidth}\resizebox{\textwidth}{!}{\import{heatmaps}{hm_ivb_loop_difftune.pgf}}\end{subfigure}\par\bigskip

\begin{subfigure}[t]{0.33\textwidth}\resizebox{\textwidth}{!}{\import{heatmaps}{hm_ivb_loop_cqa.pgf}}\end{subfigure}~
\begin{subfigure}[t]{0.33\textwidth}\resizebox{\textwidth}{!}{\import{heatmaps}{hm_ivb_loop_baseline.pgf}}\end{subfigure}
\caption{Heatmaps for \bhivel for basic blocks with a measured throughput of less than 10 cycles/iteration on Ivy Bridge}
\end{figure}
\vfill

\end{document}